\documentclass[sigconf]{acmart}

\usepackage{amsmath}
\usepackage[capitalise,noabbrev]{cleveref}
\usepackage{subfigure}

\copyrightyear{2022}
\acmYear{2022}
\setcopyright{licensedothergov}
\acmConference[WNS3 2022]{Proceedings of the WNS3 2022}{June 22--23, 2022}{Virtual Event, USA}
\acmBooktitle{Proceedings of the WNS3 2022 (WNS3 2022), June 22--23, 2022, Virtual Event, USA}
\acmPrice{15.00}
\acmDOI{10.1145/3532577.3532607}
\acmISBN{978-1-4503-9651-6/22/06}

\begin{document}

\title{Machine Learning Based Propagation Loss Module for Enabling Digital Twins of Wireless Networks in ns-3}

\author{Eduardo Nuno Almeida}
\affiliation{%
    \institution{INESC TEC and Faculdade de Engenharia, Universidade do Porto}
    \city{Porto}
    \country{Portugal}
}
\email{eduardo.n.almeida@inesctec.pt}

\author{Mohammed Rushad}
\author{Sumanth Reddy Kota}
\affiliation{%
    \institution{Wireless Information Networking Group (WiNG), National Institute of Technology Karnataka, Surathkal}
    \city{Mangalore}
    \country{India}
}
\email{mohammedrushad.181co232@nitk.edu.in}
\email{sumanthreddy.181co225@nitk.edu.in}

\author{Akshat Nambiar}
\author{Hardik L. Harti}
\affiliation{%
    \institution{Wireless Information Networking Group (WiNG), National Institute of Technology Karnataka, Surathkal}
    \city{Mangalore}
    \country{India}
}
\email{aks.181co204@nitk.edu.in}
\email{hardikharti.181co220@nitk.edu.in}

\author{Chinmay Gupta}
\author{Danish Waseem}
\affiliation{%
    \institution{Wireless Information Networking Group (WiNG), National Institute of Technology Karnataka, Surathkal}
    \city{Mangalore}
    \country{India}
}
\email{chinmay.181co215@nitk.edu.in}
\email{dan.181co116@nitk.edu.in}

\author{Gonçalo Santos}
\author{Helder Fontes}
\author{Rui Campos}
\affiliation{%
    \institution{INESC TEC and Faculdade de Engenharia, Universidade do Porto}
    \city{Porto}
    \country{Portugal}
}
\email{goncalo.r.santos@inesctec.pt}
\email{helder.m.fontes@inesctec.pt}
\email{rui.l.campos@inesctec.pt}

\author{Mohit P. Tahiliani}
\affiliation{%
    \institution{Wireless Information Networking Group (WiNG), National Institute of Technology Karnataka, Surathkal}
    \city{Mangalore}
    \country{India}
}
\email{tahiliani@nitk.edu.in}

\begin{abstract}
The creation of digital twins of experimental testbeds allows the validation of novel wireless networking solutions and the evaluation of their performance in realistic conditions, without the cost, complexity and limited availability of experimental testbeds. Current trace-based simulation approaches for ns-3 enable the repetition and reproduction of the same exact conditions observed in past experiments. However, they are limited by the fact that the simulation setup must exactly match the original experimental setup, including the network topology, the mobility patterns and the number of network nodes. In this paper, we propose the Machine Learning based Propagation Loss (MLPL) module for ns-3. Based on network traces collected in an experimental testbed, the MLPL module estimates the propagation loss as the sum of a deterministic path loss and a stochastic fast-fading loss. The MLPL module is validated with unit tests. Moreover, we test the MLPL module with real network traces, and compare the results obtained with existing propagation loss models in ns-3 and real experimental results. The results obtained show that the MLPL module can accurately predict the propagation loss observed in a real environment and reproduce the experimental conditions of a given testbed, enabling the creation of digital twins of wireless network environments in ns-3.
\end{abstract}

%
%


\maketitle

\renewcommand{\shortauthors}{E. N. Almeida, M. Rushad, S. R. Kota, A. Nambiar, H. L. Harti, C. Gupta, D. Waseem, et al.}

\section{Introduction}

To develop novel solutions for next-generation wireless networks, researchers need to evaluate the performance of their solutions in realistic scenarios. Experimental testbeds are often the preferred approach since they consider all natural physical phenomena characterising the environment. However, the cost and complexity of setting up experimental testbeds, as well as their limited availability, bring up challenges to obtaining results with statistical confidence. Even if the testbed is available, the variable conditions of the environment, including external phenomena, limit the ability to accurately repeat and reproduce experiments, which may lead to results that differ from those originally obtained.

Network simulators, including ns-3 \cite{henderson2008network}, provide the ability to easily repeat and reproduce experiments, using the same exact conditions specified in the simulation. The relative simplicity of setting up a simulation and the availability of propagation loss models provide sufficiently accurate estimations of the solution's performance. However, in extreme scenarios with dynamic environment conditions, the results obtained in simulation may significantly differ from the experimental testbed. Existing models are generic and do not capture the specific characteristics of a given physical environment.

In order to overcome this problem, trace-based simulation approaches have been proposed to replicate the experimental conditions of the environment in simulation, working at the application layer \cite{agrawal2016trace, owezarski2004trace}, at the physical layer \cite{fontes2017trace, fontes2018improving, lamela2019repeatable, cruz2020reproduction}, or even considering a combination of the medium access control and physical layers \cite{lamela2021reproducible}. In particular, the solutions proposed for the physical layer enable ns-3 to accurately repeat and reproduce the physical conditions of an environment, using network traces collected in past experiments. Despite the increased accuracy and realism of these solutions, the simulation setup must match exactly the experimental setup, since the data contained in the network traces, such as the Signal-to-Noise Ratio (SNR) of the received packets, are directly applied to the packets generated in ns-3 for the same exact node, coordinates and time instant. The use of Machine Learning (ML) techniques to predict the propagation loss in urban scenarios has been explored in \cite{zhang2019path, cabral2019machine, popescu2006ann, popescu2006comparison}. Despite the interesting results, these solutions are not integrated in network simulators. Additionally, they only estimate the path loss component of the propagation loss, and do not consider the stochastic fast-fading component affecting the radio signal propagation.

In this paper, we propose the \textbf{ML-based Propagation Loss (MLPL) module} for ns-3. The MLPL module is able to reproduce, in simulation, the experimental conditions measured in a physical environment, with the flexibility of allowing any network topology, mobility pattern, offered traffic and duration of the simulation. Based on network traces collected in an experimental testbed, the ML model is able to estimate the radio propagation loss between two nodes at any position and time instant. Therefore, the MLPL module enables the creation of a digital twin of the original wireless network environment, which allows the validation of novel solutions and the evaluation of their performance in realistic conditions in ns-3, provided that the network traces are representative of the dynamic conditions of the environment.

The MLPL module is built upon a supervised learning model, which is trained with network traces collected in the experimental testbed (the physical twin). It is composed of two sub-models: 1) the path loss model and 2) the fast-fading model. The path loss model is deterministic and estimates the path loss for a given distance between two nodes. The fast-fading component is modelled as a stochastic process that generates random samples according to the Probability Distribution Function (PDF) characterising the phenomena. This distribution may be defined as an empirical PDF or as a well-known statistical distribution -- e.g. Normal, Rayleigh and Rician. Therefore, the final propagation loss suffered by the radio signal travelling between two nodes is calculated as the sum of the path loss and the fast-fading components. This value is then used by ns-3 to calculate the received power at the destination node, considering the transmission power and the antenna gains of both nodes.

The MLPL module is validated with unit tests. Moreover, we demonstrate the usage of the module in a specific experimental environment, using the corresponding network traces, and compare the results obtained with the MLPL module, the existing propagation loss models in ns-3 and the real experimental results.

The rest of this paper is organised as follows. \cref{section:mlpl-module} explains the proposed MLPL module and the unit tests. \cref{section:validation} presents the validation of the MLPL module. \cref{section:conclusions} draws the conclusions and future work.

\section{MLPL Module} \label{section:mlpl-module}

This section provides the details of the MLPL module and the unit tests developed to validate its functionality.

\subsection{Overview}

The MLPL module is designed to allow dynamic generation of scenarios based on previously collected wireless network traces. A custom \texttt{PropagationLossModel}, similar in structure to existing propagation loss models in ns-3, has been designed to ease the integration of the MLPL module with general use in the simulator. The module has two main components: 1) the \texttt{MlPropagationLossModel} in ns-3 and 2) ns3-ai scripts to interact with external ML libraries. This approach enables the use of any ML Python library -- e.g., Tensorflow, PyTorch, SciPy -- without integrating them directly in ns-3.

\begin{figure}
    \centering
    \includegraphics[width=1\linewidth]{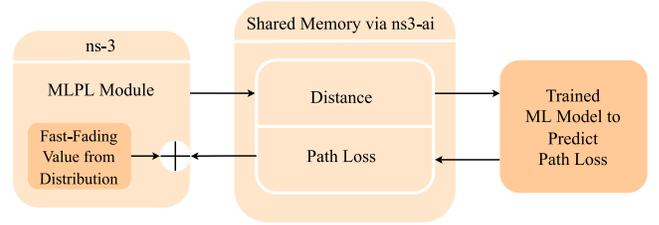}
    \caption{The MLPL Module and its Interactions with the ns3-ai and the Deterministic Path Loss to Compute the Total Propagation Loss Value}
    \label{fig:mlpl-module}
\end{figure}

The \texttt{MlPropagationLossModel} uses two values in determining propagation loss: 1) a deterministic path loss component and 2) a stochastic fast-fading component. When a new propagation loss value is requested from the MLPL module, given the distance between the nodes, the path loss component is predicted by the ML model after which a pseudo-random value from the fast-fading distribution is added to get the total propagation loss value, as depicted in \cref{fig:mlpl-module}. This result is then combined with the transmission power and antenna gains to get the actual received power. On integrating the MLPL module with ns-3, it is possible to run network simulations in conditions similar to a given recorded experiment, even if the network topology differs from the experiment, providing more realistic results than a pure simulation. However, before the MLPL module can be used in ns-3, the ML models that will be used by ns3-ai have to be prepared. These ML models will be used to predict the path loss values for the incoming requests by ns-3. For the ML models, the user can use the already available Python scripts to generate trained ML models or choose to use their own. These Python scripts preprocess the given trace data used to train the ML models and also generate the fast fading distribution values, which will be used in the \texttt{MlPropagationLossModel} component. The code for the MLPL module has been made openly available on GitLab \cite{gitlabRepo}.

\subsection{Design and Implementation}

\begin{figure}
    \centering
    \includegraphics[width=0.9\linewidth]{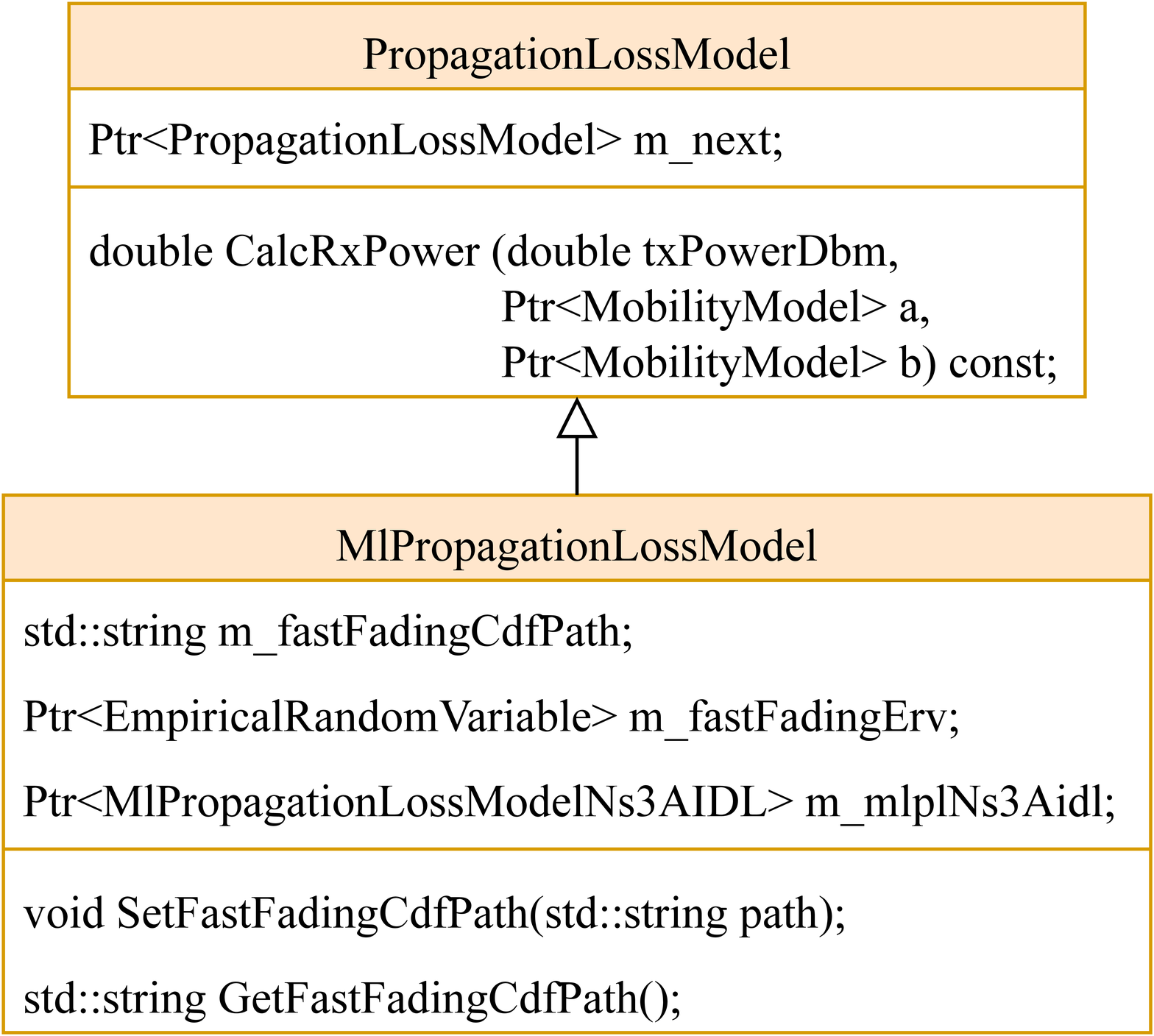}
    \caption{Class Diagram for MlPropagationLossModel ns-3 Implementation}
    \label{fig:mlpl-class-diagram}
\end{figure}

The MLPL module is implemented in ns-3 in a new class, named \texttt{MlPropagationLossModel}, which is inherited from the \linebreak \texttt{PropagationLossModel} base class as shown in \cref{fig:mlpl-class-diagram}. The \linebreak method \texttt{SetFastFadingCdfPath()} contains the implementation in the new class for creating the fast-fading distribution through the \texttt{EmpiricalRandomVariable} instance, \texttt{m\_fastFadingErv}, to \linebreak use along with the ML model while calculating the propagation loss. Further, the variable \texttt{m\_mlplNs3Aidl} is used as a pointer to the shared memory for interacting with the Python process running the ML model. In what follows, we refer to the two propagation loss components considered by the MLPL module.

\subsubsection{Deterministic Path Loss Component}

\begin{figure*}
    \centering
    \includegraphics[width=0.75\linewidth]{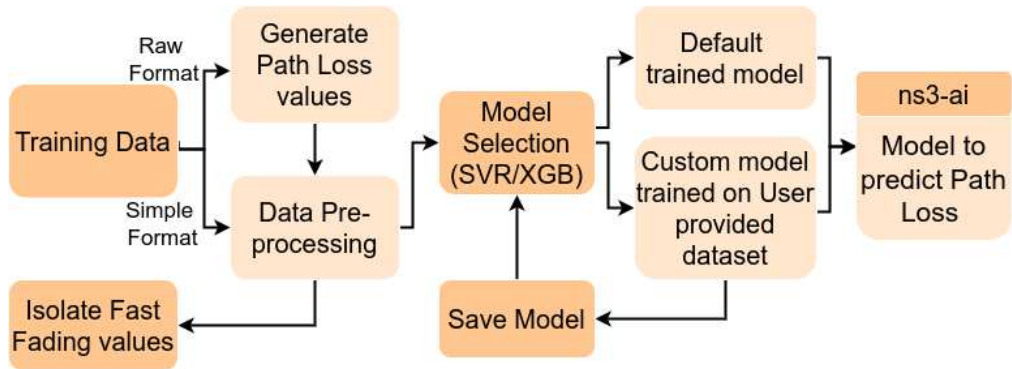}
    \caption{ML Model Generation for Predicting Path Loss}
    \label{fig:mlpl-flow-diagram}
\end{figure*}

The deterministic path loss component calculates the signal power loss that occurs as the distance increases. Two supervised learning models were trained on this data, a gradient boosted decision tree (XGBoost) and Support Vector Regression (SVR), with the objective of predicting the average path loss value as a function of the distance between the transmitter and the receiver. The user can use a trained model which is available by default in the MLPL module or use a custom trained model. For the convenience of the users, the MLPL module provides a set of scripts that can be used to generate custom trained models. \cref{fig:mlpl-flow-diagram} illustrates the process of generating the ML model.

\subsubsection{Stochastic Fast-Fading Component}

In order to model the fast-fading component of the trace data, each trace's propagation loss value is subtracted by the average propagation loss value per metre. Normal, Rician and Rayleigh distributions are fitted to the data, the probability density function is calculated and the sum of the squared error is measured. The distribution with the minimum sum of squared error is selected as the fast-fading model to be used. The distribution fitting is done using the SciPy distribution fit function, which estimates the distribution parameters that best adapt to the provided data. This generated distribution is saved as a collection of fast fading loss -- Cumulative Distribution Probability (CDF) pairs.

Since the fast-fading component is stochastic and dependent on the simulation seed, we chose to generate these samples in ns-3 using a Random Number Generator (RNG) rather than through ns3-ai. This allows the RNG to be controlled by the same ns-3 seed and makes the simulations repeatable. When an instance of the \texttt{MlPropagationLossModel} is created, the pairs  are read using the \texttt{SetFastFadingCdfPath()} function, which configures an instance of the \texttt{EmpiricalRandomVariable} according to  this distribution. When a fast-fading loss value is requested from the \linebreak \texttt{EmpiricalRandomVariable}, a pseudo-random sample is chosen from the generated distribution.

\subsection{ML Trace-Based Dataset}

The MLPL module allows the user to use two different formats for the dataset as explained below.

\subsubsection{Simple Data Format}

The user can provide a simple dataset consisting of the distances and their respective path loss values. We only use the path loss component to train the ML model, so it is fundamental to isolate this component from the fast-fading. This leads to a conjecture that the fast-fading component can be modelled as a Gaussian distribution with a mean of zero, as we are assuming that the average of the complete propagation loss values will counterbalance the fast-fading aspect. \cref{fig:mlpl-flow-diagram} shows the flow of the model when a simple data format is used.

\subsubsection{Raw Data Format}

The ML trace-based dataset is a raw \linebreak dataset which is composed of the transmission power, the SNR, the sender / receiver node coordinates, and other configuration details of the dataset such as the antenna gain and channel frequency. To make the prediction model independent of the transmission power, the collected SNR value is not used directly as training data. Instead, a pre-processing step is performed. The distance is calculated using the nodes' coordinates noted in its simulation environment, which is reproducing the waypoints of the real experiment. \cref{fig:mlpl-flow-diagram} also shows the flow of the model when raw data format is used.

\subsection{Preliminary Validation with Unit Tests}

To validate the functionality of the MLPL module, a new test suite named \texttt{MlPropagationLossModelTest} has been developed. This test suite contains one fundamental test comprising four \linebreak \texttt{TestVectors} for checking distance -- path loss values.

This test suite is primarily targeted at validating the working of the shared memory via ns3-ai in the \texttt{MlPropagationLossModel} for expected prediction of path loss values using the trained ML model. Additionally, it also validates the creation of the \linebreak \texttt{EmpiricalRandomVariable} using the CDF distribution for obtaining pseudo-random fast-fading values.

\section{Validation of the MLPL Module} \label{section:validation}

We validate the MLPL module by following a two-step approach. Firstly, we validate the performance of the ML models by comparing the predicted path loss values in dB to those obtained from real experiments conducted on a live wireless network, and comparing these results with those obtained with Friis and Log-distance propagation loss models, available in ns-3. Subsequently, we use the ML models in ns-3 and validate the effectiveness of the MLPL module by comparing the goodput obtained in a specific scenario when using  existing propagation loss models in ns-3, such as Friis and Log-distance. At every step of validation, we discuss the experimental setup, parameters, scenarios and analyse the results.

\subsection{Accuracy of the ML Models}

In this section, we validate the accuracy of the ML models in several experimental scenarios by comparing the path loss values obtained with the ML models to those observed in the real experiments.

\subsubsection{Experimental Set-Up}
\label{experimental-setup}

\begin{figure*}
    \centering
    \includegraphics[width=0.9\linewidth]{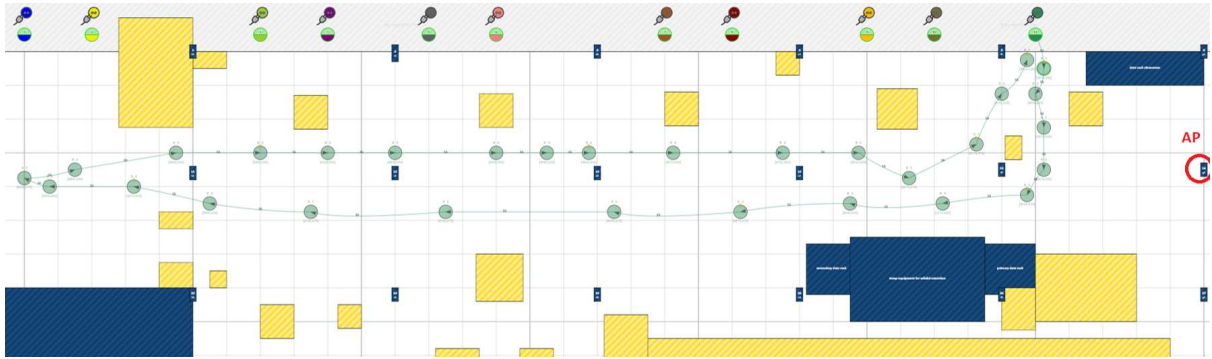}
    \caption{Trajectory of the Mobile Node in Environment}
    \label{fig:mobile-nodes-trajectory}
\end{figure*}

The set-up is based on an experiment performed in the SIMBED project \cite{simbedDataset}, consisting of data collected using Fed4FIRE+ testbeds \cite{fed4firePlus}, in a warehouse environment. We consider two nodes -- one fixed and one mobile -- which provides us with continuous distance values as the mobile node moves through the environment. \cref{fig:mobile-nodes-trajectory} presents the fixed node as the "AP", and also the trajectory of the mobile node. The collected traces in the project registered multiple metrics, such as the distance, the SNR and the goodput. Each experiment consists of multiple runs, varying the node's transmission power from 0 dBm to 12 dBm.

\subsubsection{Experimental Scenarios} \label{experimental-scenarios}

The ML models are trained on either complete or partial data from an experiment run and evaluated on the respective data as explained in the scenarios below. These ML models are measured against other path loss models like Friis and various Log-distance models by comparing the propagation loss predicted by each of them with the real measurements. The following are the scenarios considered:

\begin{itemize}

    \item \textbf{Extrapolation Scenario:} In this scenario, the ML models are trained on shorter distances and tested on longer distances, to evaluate its extrapolation capabilities. Only data within 10 m between the nodes is available for training, the models are evaluated on all other data.

    \item \textbf{Interpolation Scenario:} In this scenario, the distances are split into bins from which the models could learn. The training data consists of data from distances below 5 m, between 10 and 15 m, and finally above 20 m, effectively training on data with knowledge gaps.

    \item \textbf{Full-Set Scenario:} In this final scenario, the models are trained on all available data from a particular run of the experiment.

\end{itemize}

The training data originated from run \texttt{08022019\_11.04.35} \cite{simbedDataset}, which is characterised by a transmission power of 1 dBm. The test data comes from run \texttt{07022019\_02.49.27} \cite{simbedDataset}, where a transmission power of 7 dBm was used. Both these data runs share the same channel centre frequency (5220 MHz) and bandwidth (20 MHz). A gain of -7 dBi was used for each antenna. The gain of the antennas is represented as a negative value since signal attenuators of 10 dB were used in-line with the 3 dBi antennas, to limit the signal's range in the warehouse.

\subsubsection{Results and Analysis}

The results obtained for the three experimental scenarios are analysed in the following subsections.

\paragraph{Extrapolation Scenario}

As mentioned in \cref{experimental-scenarios}, this scenario captures only the data under 10 metres for training. The behaviour of the ML models can be seen in \cref{fig:extrapolation-results} given below.

In \cref{fig:extrapolation-path-loss}, the thick blue dashed line represents the median values, while the shadow boundaries represent the 25th and 75th percentiles. When the fast-fading component is removed, the 25th and 75th percentiles are not shown on the plots of Friis and Log-distance models since these models are deterministic. While predicting propagation loss for known distance values (< 10m), the SVR and XGBoost predictions follow the real data values closely. However, when dealing with unseen data -- i.e., data not included in the training set -- both ML trace-based models predict virtually a constant value. As expected, the Friis and Log-distance models' behaviour does not change.

\cref{fig:extrapolation-percentile} shows us the absolute difference between the 25th, 50th (median) and 75th percentiles for each model and the real data. The perfect scenario would be a horizontal line through 0 on the Y-axis, where both models' percentiles have the same value. Here, we see a similar situation, where the difference curve for unseen values follows the real data one, as the predicted values are constant.

The boxplot in \cref{fig:extrapolation-box-plot} represents the propagation loss distribution for each model, highlighting the median, the 25th and the 75th percentiles. The whiskers depict the full distribution, excluding some points considered to be outliers, shown individually. The points are considered outliers if they fall outside of the area defined by multiplying the interquartile range, 75th Percentile -- 25th Percentile, by 1.5, subtracting it to the 25th Percentile and adding it to the 75th Percentile. When comparing the propagation loss values distribution as shown in \cref{fig:extrapolation-box-plot}, the results are again similar, with \textbf{XGBoost providing the most accurate representation of the data}. With all the above information, it is possible to infer that the extrapolation capabilities of the ML trace-based propagation loss model are limited and other training strategies should be preferentially used.

\begin{figure}
    \centering
    \subfigure[Path Loss Prediction by Distance] {
        \includegraphics[width=1\linewidth]{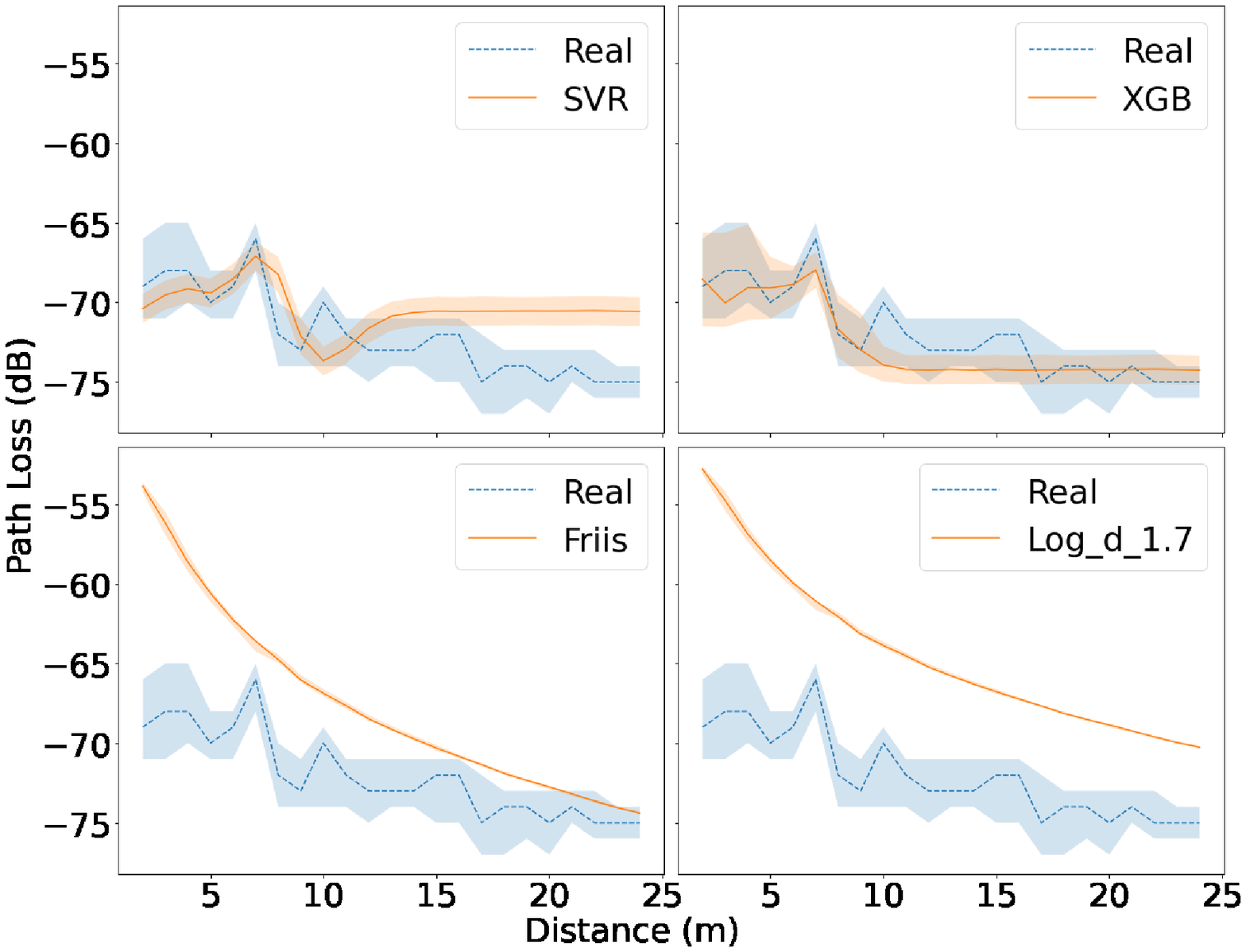}
        \label{fig:extrapolation-path-loss}
    }
    \vfil
    \subfigure[Path Loss Percentile by Distance] {
        \includegraphics[width=1\linewidth]{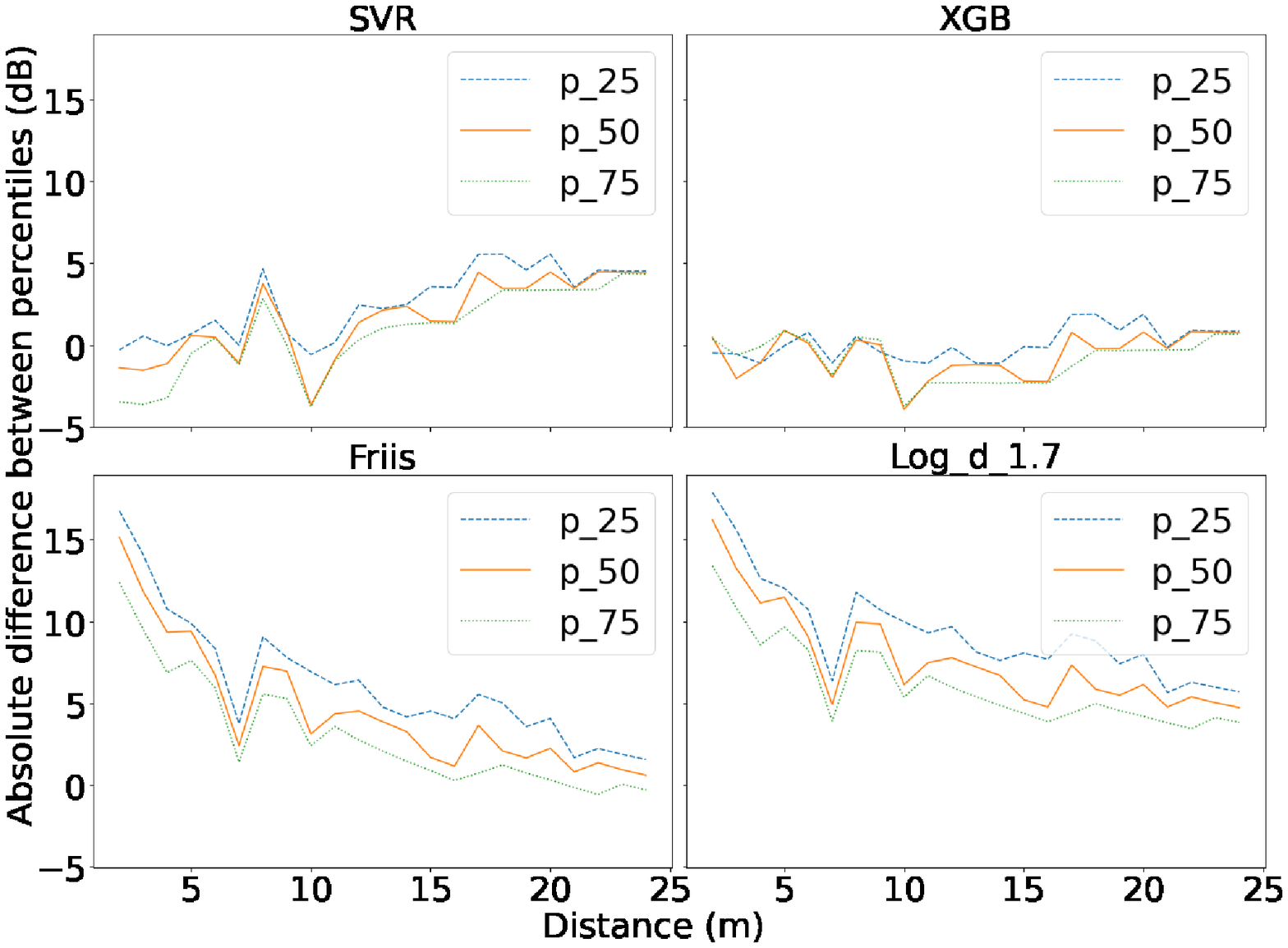}
        \label{fig:extrapolation-percentile}
    }
    \vfil
    \subfigure[Path Loss Distribution by Algorithm] {
        \includegraphics[width=1\linewidth]{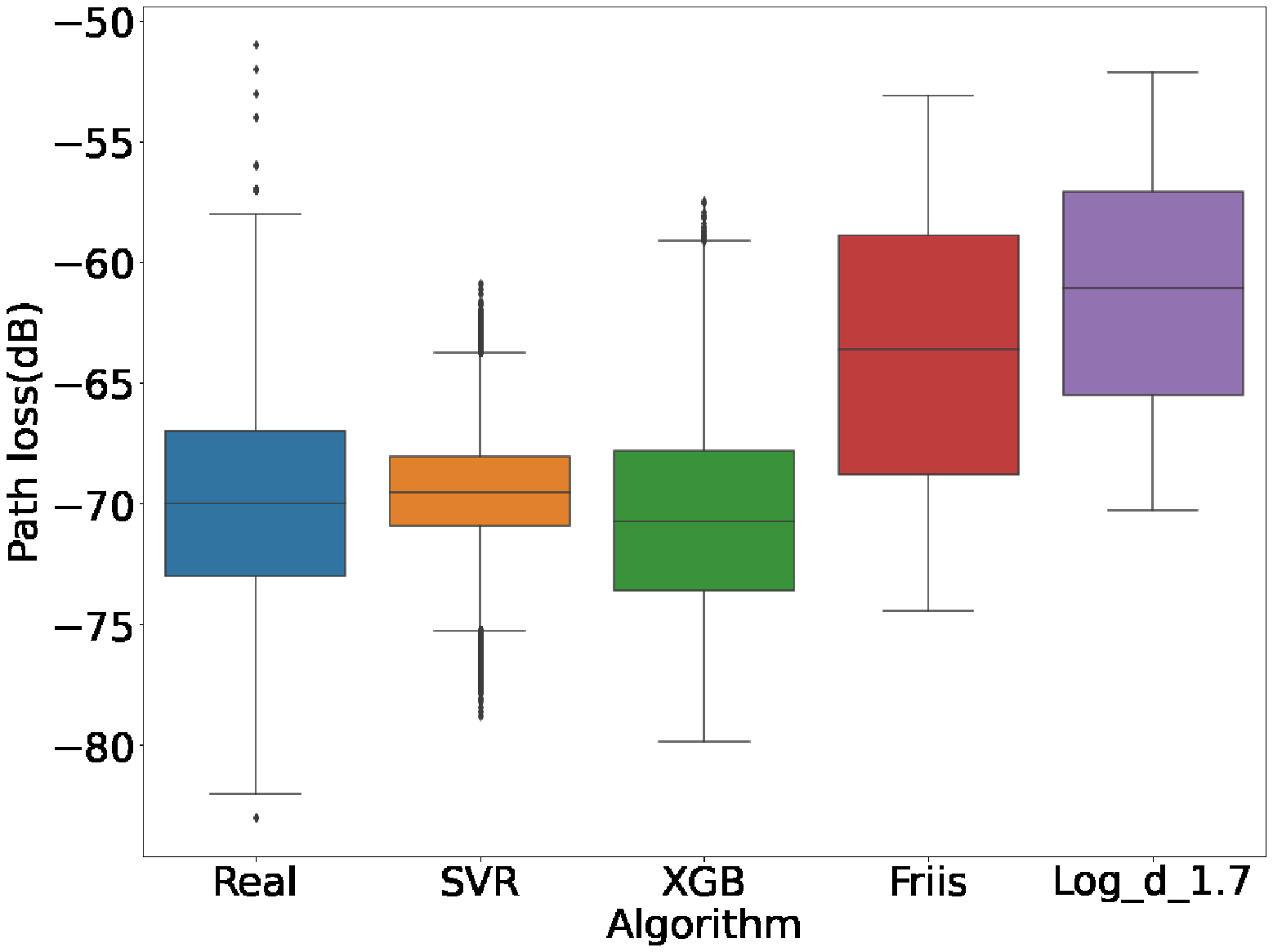}
        \label{fig:extrapolation-box-plot}
    }
    \caption{Results for the Extrapolation Scenario}
    \label{fig:extrapolation-results}
\end{figure}

\paragraph{Interpolation Scenario}

\begin{figure}
    \centering
    \subfigure[Path Loss Prediction by Distance]{
        \includegraphics[width=1\linewidth]{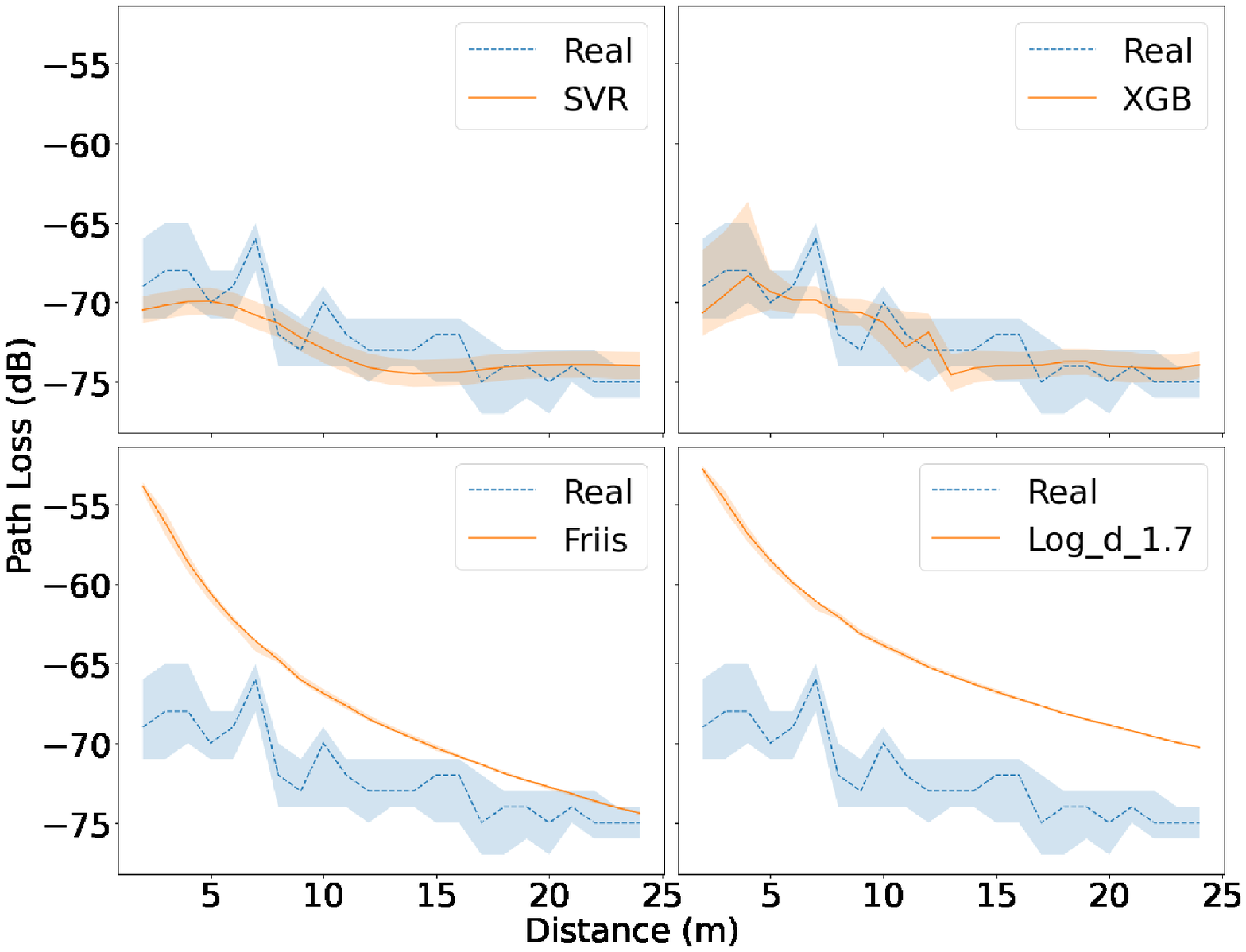}
        \label{fig:interpolation-path-loss}
    }
    \vfil
    \subfigure[Path Loss Percentile by Distance]{
        \includegraphics[width=1\linewidth]{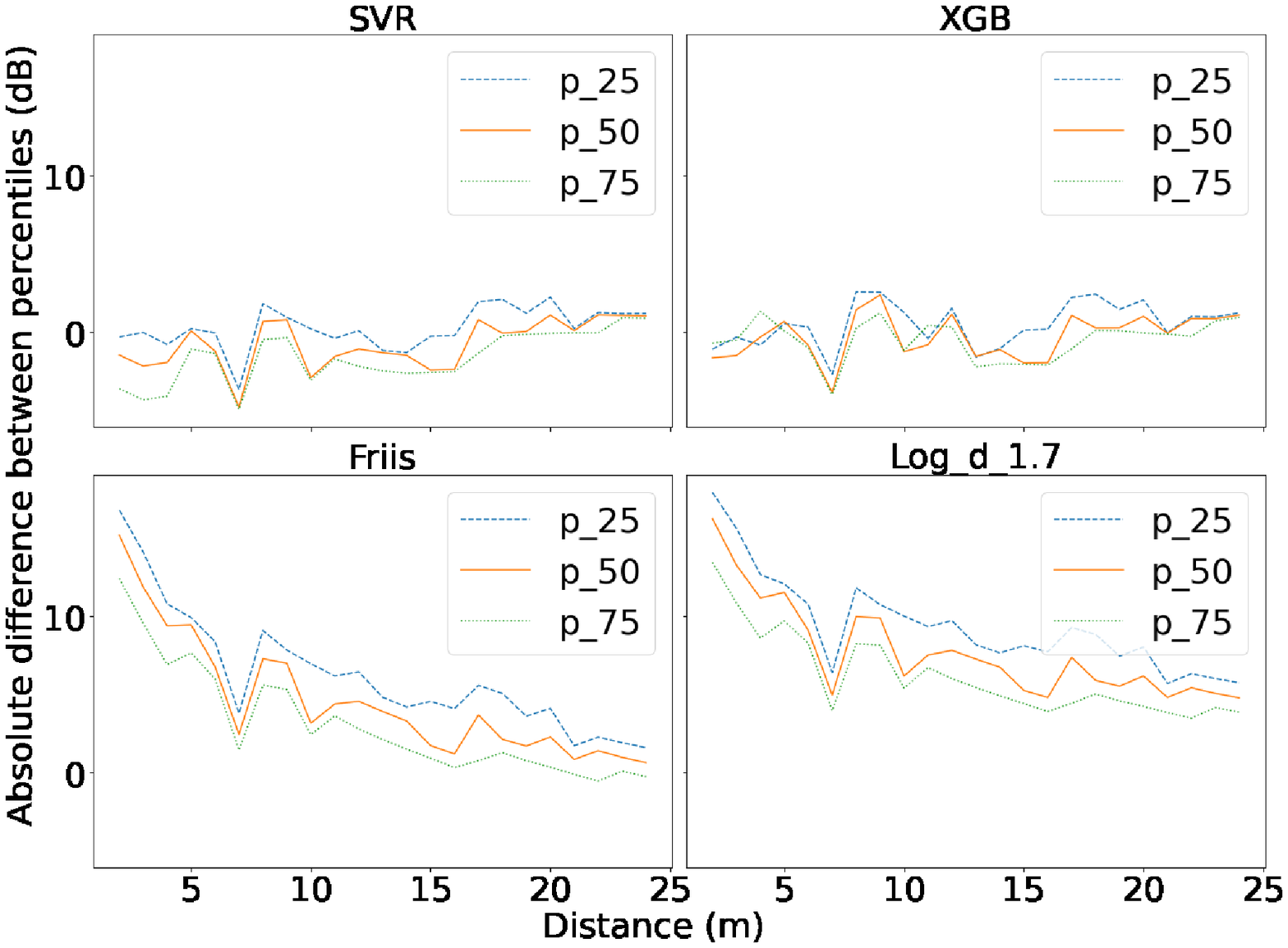}
        \label{fig:interpolation-percentile}
    }
    \vfil
    \subfigure[Path Loss Distribution by Algorithm]{
        \includegraphics[width=1\linewidth]{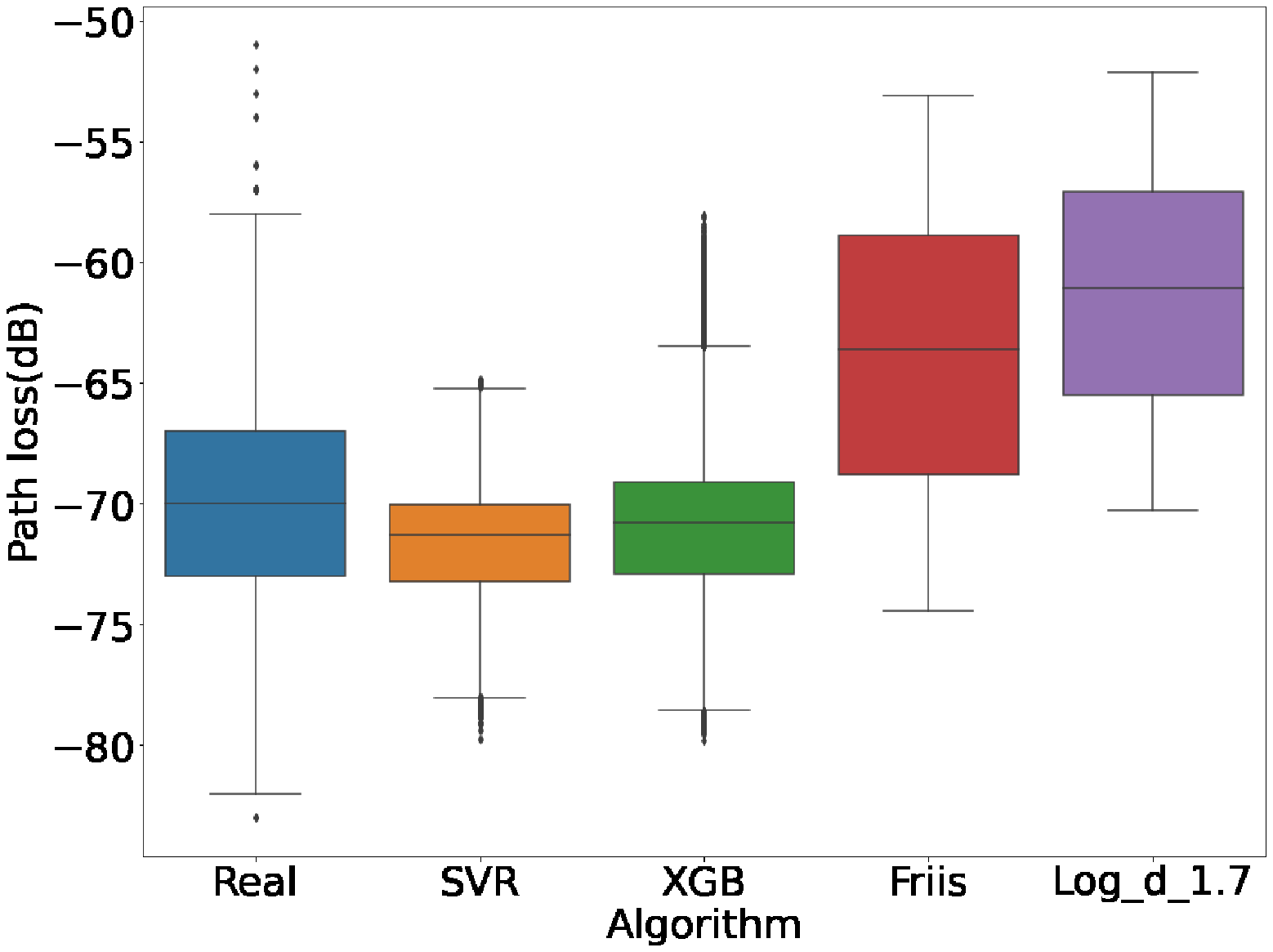}
        \label{fig:interpolation-box-plot}
    }
    \caption{Results for the Interpolation Scenario}
    \label{fig:interpolation-results}
\end{figure}

For this scenario, a different strategy is employed. The training data for the models have some gaps. It consists of data from distances shorter than 5 m, between 10 m and 15 m and larger than 20 m. The results are shown in \cref{fig:interpolation-results}.

As can be seen in \cref{fig:interpolation-path-loss}, both SVR and XGBoost curves are essentially contained in the real data shadow, except on the local spikes excluded from train data, where both models follow the overall tendency on that spot. This indicates that this approach still allows us to achieve good performance while using less data.

Looking at the absolute difference between the 25th, 50th (median) and 75th percentiles for each model and the real data in \cref{fig:interpolation-percentile}, it is possible to infer that the knowledge gaps were not a problem for the ML models, as they were able to learn from the previous and following values for all the missing values.

The propagation loss distribution for each model is similar to the extrapolation scenario, where SVR and XGBoost distributions are more identical to the real one than the Friis and Log-distance distributions, as can be observed in \cref{fig:interpolation-box-plot}.

\paragraph{Full-Set Scenario}

\begin{figure}
    \centering
    \subfigure[Path Loss Prediction by Distance]{
        \includegraphics[width=1\linewidth]{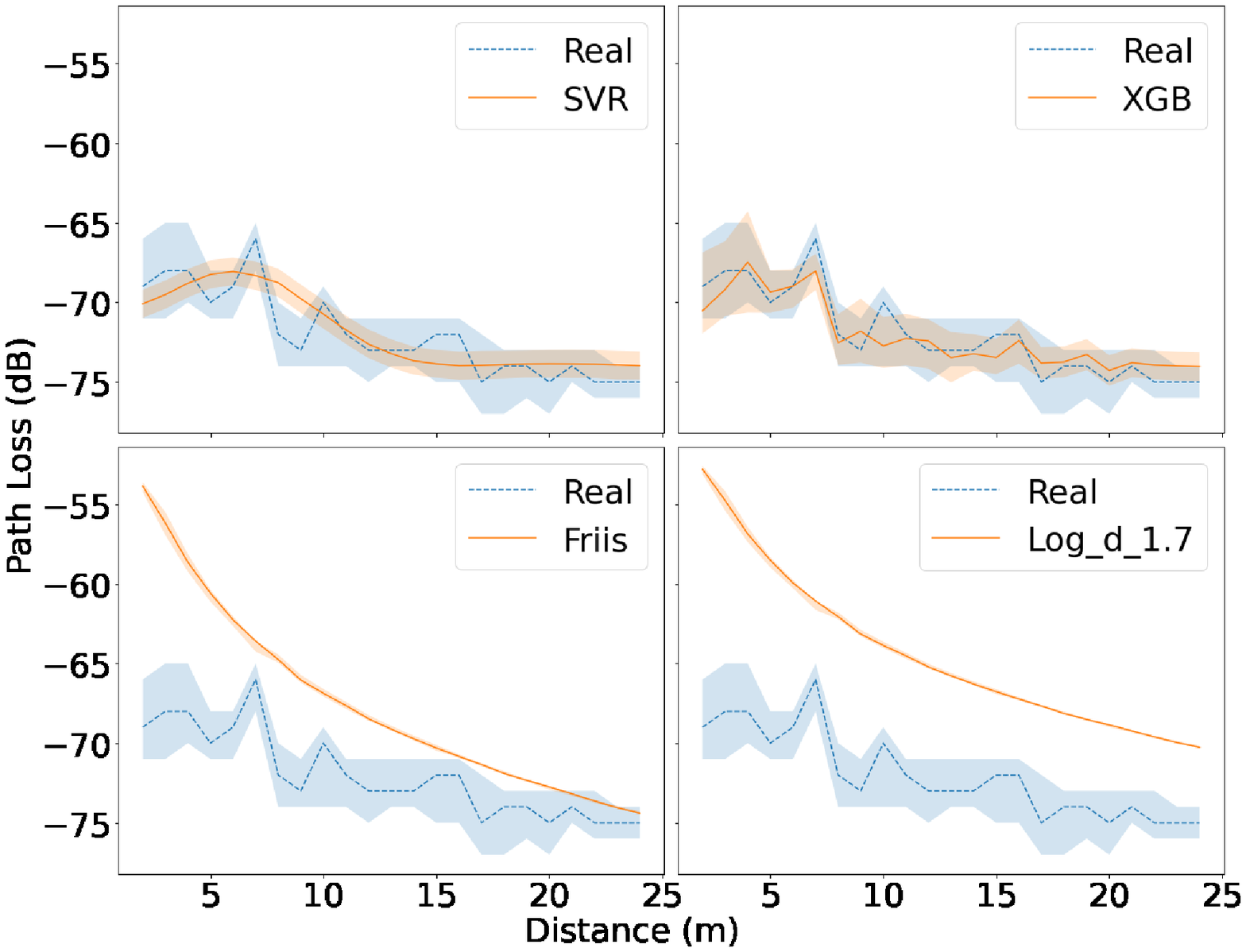}
        \label{fig:full-set-path-loss}
    }
    \vfil
    \subfigure[Path Loss Percentile by Distance]{
        \includegraphics[width=1\linewidth]{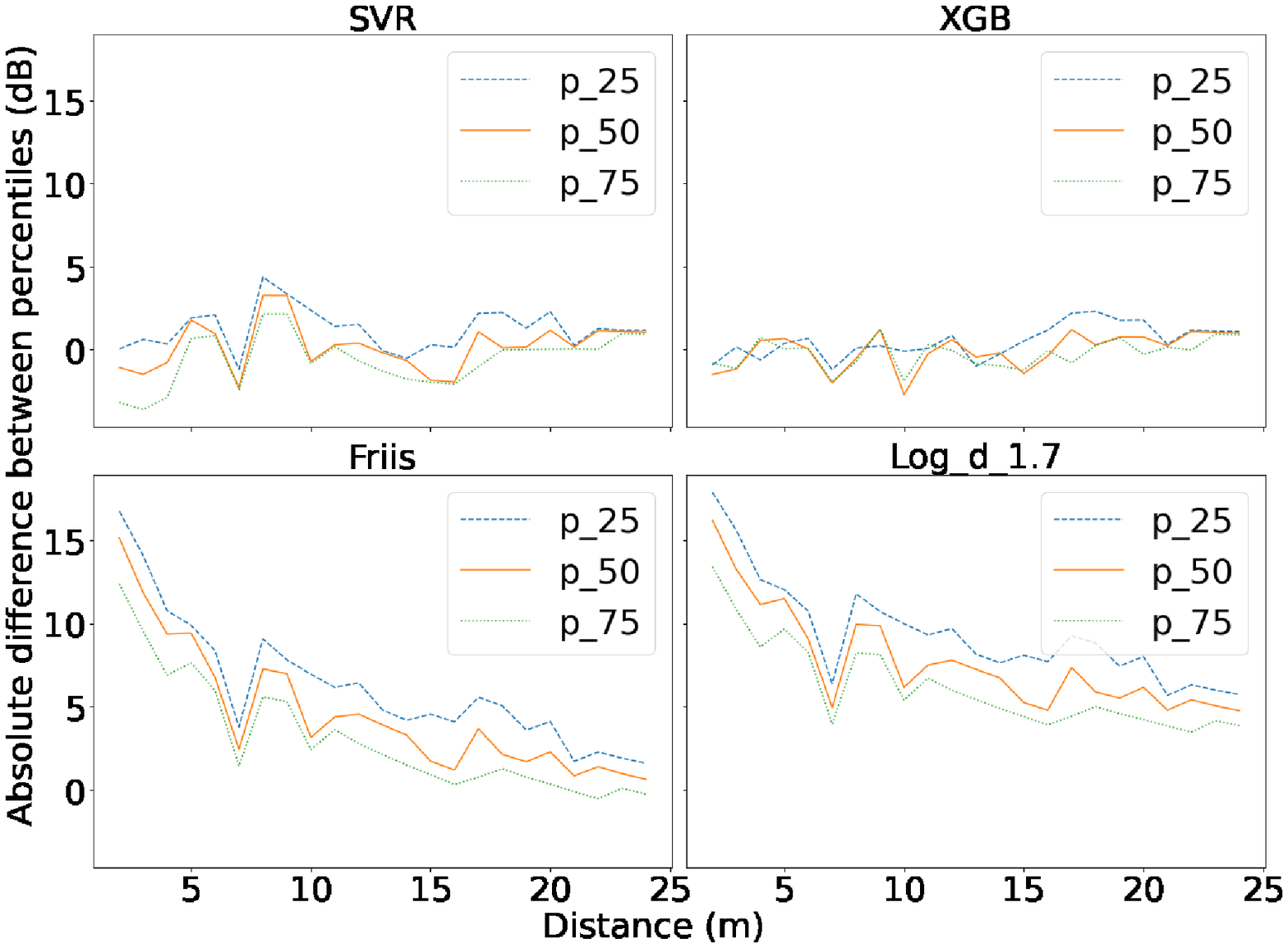}
        \label{fig:full-set-percentile}
    }
    \vfil
    \subfigure[Path Loss Distribution by Algorithm]{
        \includegraphics[width=1\linewidth]{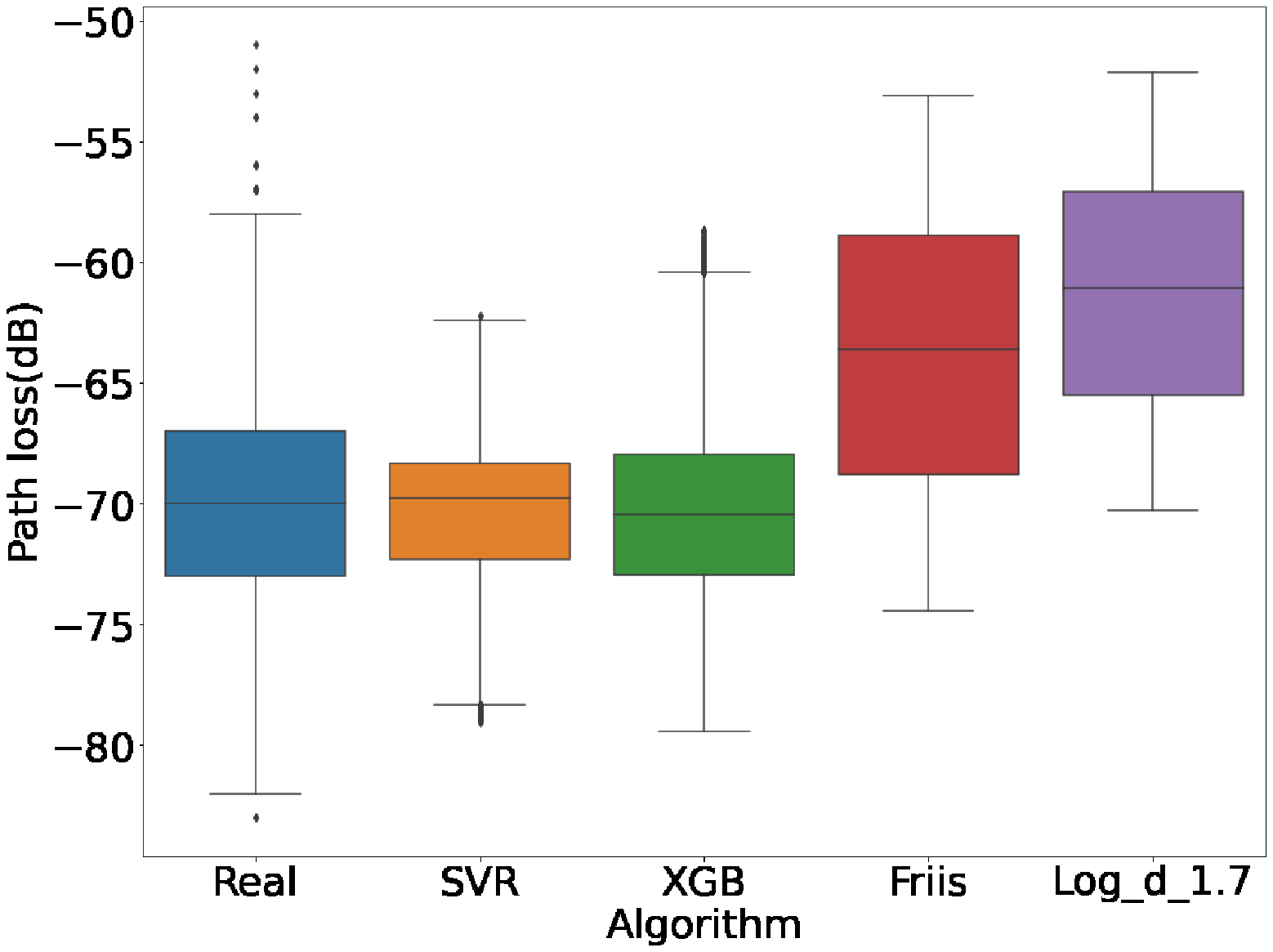}
        \label{fig:full-set-box-plot}
    }
    \caption{Results for the Full Set Scenario}
    \label{fig:full-set-results}
\end{figure}

\cref{fig:full-set-results} shows the results of the full set scenario. When all data is used to train the ML model, SVR and XGBoost have been shown to better modulate the propagation loss when compared to either Friis or multiple Log-distance path loss models.

As in \cref{fig:full-set-path-loss}, the SVR and XGBoost predictions follow the real data values closely. SVR predictions give a smoother curve in predicting the values, while XGBoost is able to replicate the local spikes along the experiment. In this context, Friis and Log-distance are too optimistic.

Comparing the absolute difference between the 25th, 50th (median), and 75th percentiles for each model and the real data, the results follow the \cref{fig:full-set-percentile} findings, where SVR and XGBoost have a smaller error. The XGBoost difference is slightly smaller than the SVR one. The Friis and Log-distance models show a maximum error around 15 dB, while ML models have errors below 5 dB.

When comparing the propagation loss distribution for each \linebreak model, in \cref{fig:full-set-box-plot}, SVR and XGBoost present similar results, being both able to accurately model the real data. The analysis of the boxplot, in conjunction with the rest of the data, shows that the ML trace-based propagation loss model can correctly learn and represent the real experiment data when trained on the full dataset.

\subsection{Effectiveness of the MLPL Module in ns-3}

In this section, we validate the effectiveness of the proposed MLPL module in ns-3. We study the goodput values obtained from simulations in ns-3 by varying the distance between nodes, and then compare these values to those obtained from the real experiments.

\subsubsection{Simulation Set-Up and Parameters}

To evaluate the impact of using the MLPL module on a specific wireless network scenario, the network performance using Friis, Log-distance and MLPL is measured and compared. The simulation set-up is in line with the experimental set-up described in \cref{experimental-setup}, involving two nodes -- one fixed and one mobile -- as shown in \cref{fig:mobile-nodes-trajectory}. The simulation is configured to use the same values of the parameters considered in the real experiment, which are described in \cref{table:Parameters}.

\begin{table}
\caption{ns-3 Simulation Parameters} \label{table:Parameters}
    \begin{tabular}{c c}
        \toprule
        \textbf{Parameter} & \textbf{Value} \\
        \midrule
        ns-3 version & 3.35 \\
        Wi-Fi standard & IEEE 802.11a \\
        Tx power & 7 dBm \\
        Antenna gains & -7 dBi \\
        Channel frequency & 5220 MHz \\
        Channel bandwidth & 20 MHz \\
        Preamble detection threshold & -90 dBm \\
        Generated traffic & 54 Mbit/s UDP constant bitrate \\
        Packet size & 1400 bytes \\
        Distance range & [2.07m, 24.09m] \\
        Simulation duration & 404 seconds \\
        \bottomrule
    \end{tabular}
\end{table}

The position of both nodes simulated in ns-3 is consistent with the data collected from real experiments. The moving node position is updated once a second, based on the resolution of the distance data. The fixed node generates a UDP flow with a constant bitrate of 54 Mbit/s to ensure that the connection is fully loaded, as the offered load is always above the link capacity. Since we use the IEEE 802.11a standard, the Minstrel rate adaptation algorithm is used to automatically adjust the data rate based on the channel condition. The traffic is generated in the simulation using the \texttt{OnOffApplication}, which is configured to be always ON and transmitting packets of size 1400 bytes. All unspecified parameters use ns-3 default values.

\subsubsection{Results and Analysis}

To study the impact of the MLPL module in simulated wireless environments, it is compared against existing propagation loss models in ns-3, such as Friis and \linebreak Log-distance. The Log-distance model is used in conjunction with \linebreak \texttt{JakesPropagationLossModel}, in order to reproduce the fast-fading component of the propagation loss. The training approach followed the "Full Set" scenario explained in \cref{experimental-scenarios}, as this model has the best performance. The propagation loss models are compared using the goodput measured on the receiving node, that is the number of bits of useful information delivered to the application layer of the receiving node per second.

\begin{figure}
    \centering
    \subfigure[Distance Over Time and Goodput over Distance for the Real Experiment Run]{
        \includegraphics[width=1\linewidth]{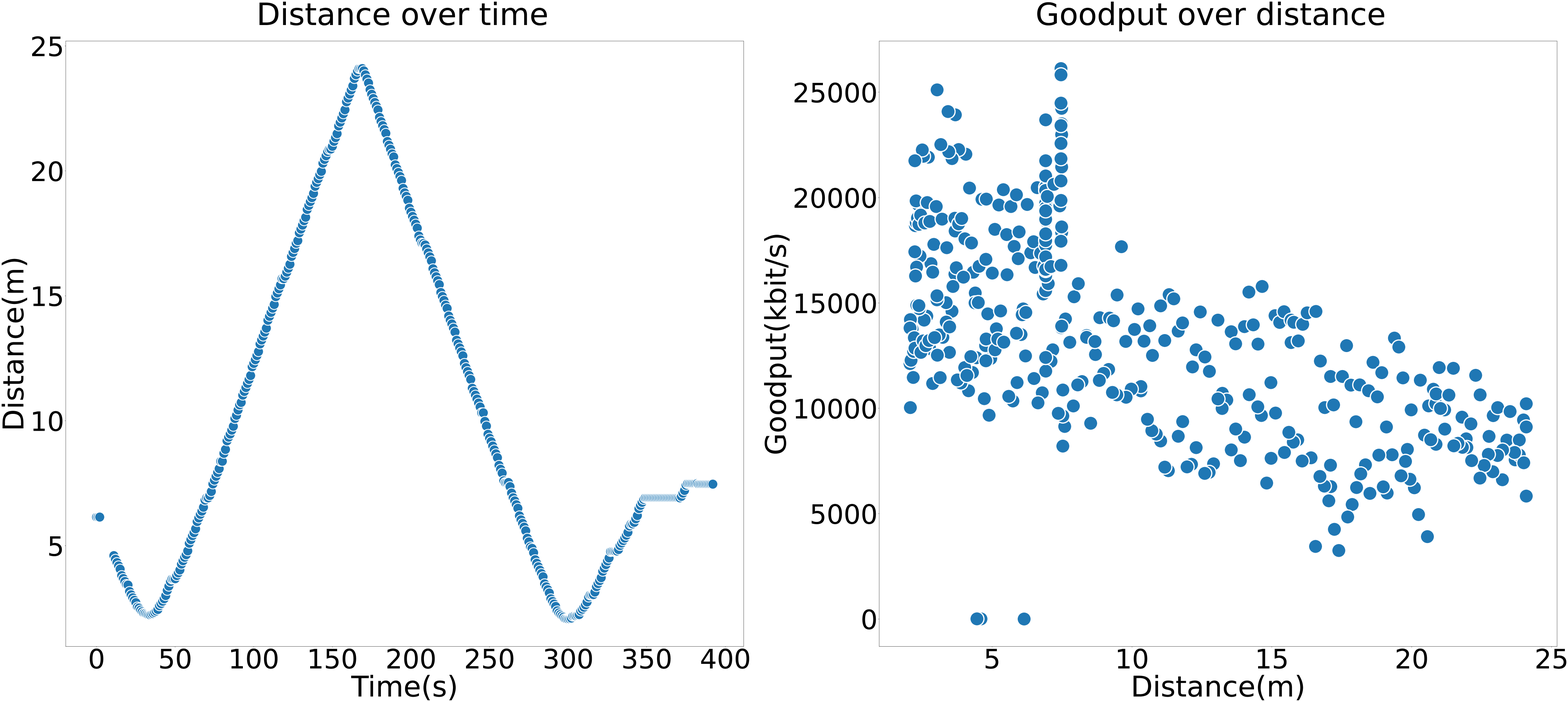}
        \label{fig:goodput-real-exp}
    }
    \vfil
    \subfigure[Goodput over Distance for the ns-3 Propagation Loss Models]{
        \includegraphics[width=1\linewidth]{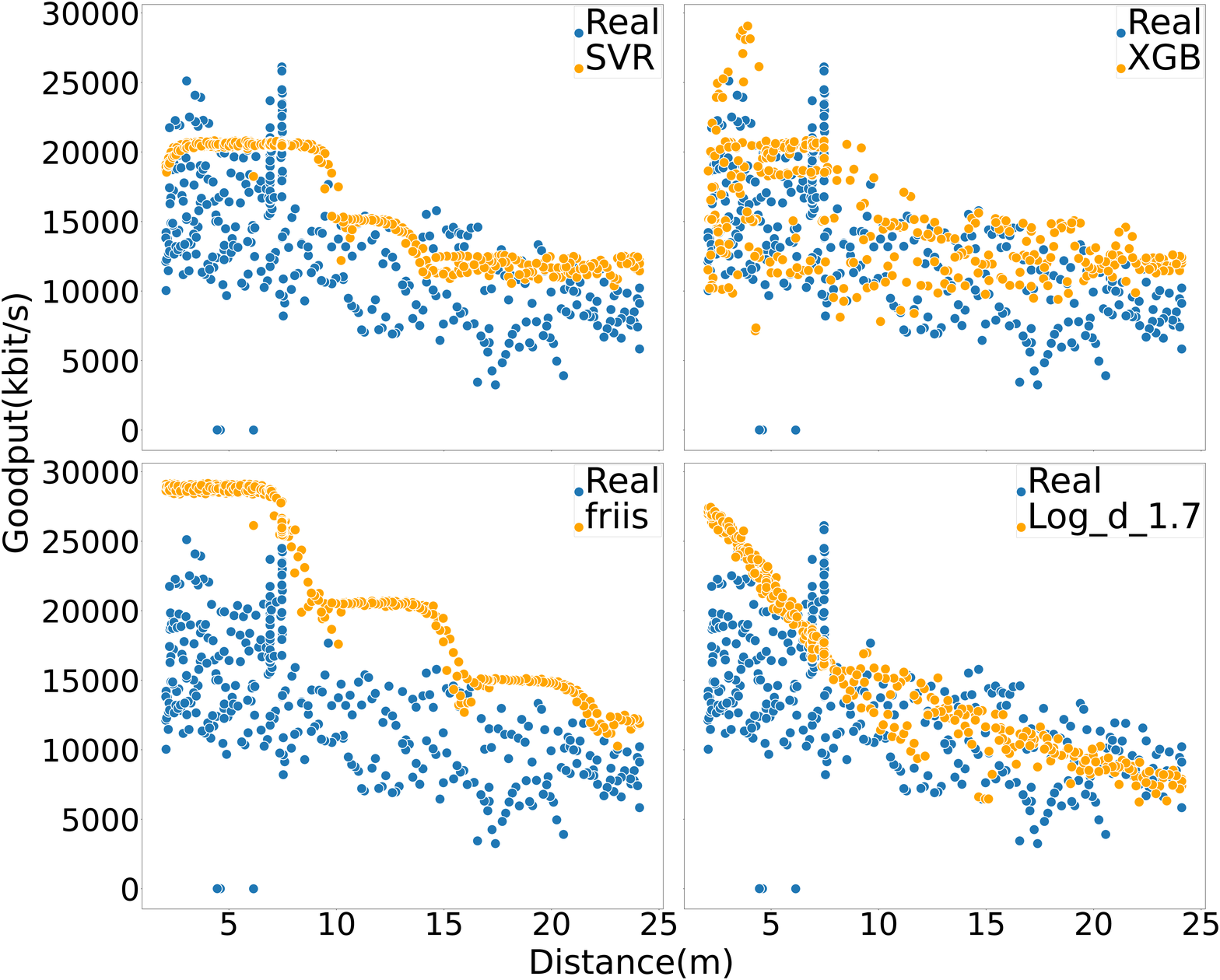}
        \label{fig:goodput-mlpl}
    }
    \caption{Results using Different Propagation Loss Models}
    \label{fig:propagation-loss-results}
\end{figure}

The results are presented in \cref{fig:propagation-loss-results}. \cref{fig:goodput-real-exp} shows the distance over time and the goodput over distance for the real experimental run. From \cref{fig:goodput-mlpl}, we observe that neither Friis nor Log-distance models are able to accurately reproduce the observed goodput, the predictions being condensed and lacking the spread of the real data. Friis is too optimistic, saturating the channel when the node distance is smaller, resulting in higher than expected goodput values during all experiments.

SVR, while not completely replicating the original results, generally is able to follow the trend in the real data. XGBoost, while being able to replicate some of the observed real values, is not able to accurately reproduce the goodput when the distance increases. However, it is the only model that can accurately reproduce the spread of the real data. Both ML models do not achieve as low goodput values as the real experiment, however, for higher values they tend to be more accurate. This data indicates that both ML models lead to more accurate representations of the real-world scenario than the pure simulation models.

\section{Conclusions and Future Work} \label{section:conclusions}

In this paper, we proposed the ML-based Propagation Loss (MLPL) module for ns-3. The MLPL module reproduces, in simulation, the experimental conditions of a physical environment, with the flexibility of allowing any network topology, offered traffic and duration of the simulation. The MLPL module provides the ability to train a supervised ML model to learn and predict deterministic path losses and the stochastic fast-fading losses using network traces collected in the environment. Therefore, the MLPL enables the creation of accurate digital twins of experimental wireless network environments in ns-3. The MLPL module was validated with unit tests. Further, we validated its effectiveness by comparing the results obtained with propagation loss models in ns-3 and experimental results. The results obtained show that the MLPL module proved to be more accurate than their counterparts when estimating the propagation loss and goodput in scenarios replicating a real experiment. As future work, we plan to submit the MLPL module to the ns-3 App Store. Additionally, we plan to improve the accuracy of the ML models to consider more parameters collected in the experimental environment.
\balance

\begin{acks}
This article is a result of the project "DECARBONIZE – DEvelopment of strategies and policies based on energy and non-energy applications towards CARBON neutrality via digitalization for citIZEns and society" (NORTE-01-0145-FEDER-000065), supported by Norte Portugal Regional Operational Programme (NORTE 2020), under the PORTUGAL 2020 Partnership Agreement, through the European Regional Development Fund (ERDF).
\end{acks}

\bibliographystyle{ACM-Reference-Format}
\bibliography{bibliography}

\end{document}